\documentclass[aps,prl,reprint,superscriptaddress,nofootinbib,floatfix]{revtex4-2}
\usepackage{amsmath,amssymb,bm,graphicx,booktabs}
\usepackage{xcolor}
\usepackage{tikz}
\usepackage{pgfplots}
\pgfplotsset{compat=1.18}
\usetikzlibrary{decorations.pathmorphing,arrows.meta,calc,shadows}
\usepackage{orcidlink}
\usepackage{hyperref}
\hypersetup{
	colorlinks=true,
	linkcolor=blue,
	filecolor=magenta,      
	urlcolor=blue,
	citecolor=magenta
}

\definecolor{normc}{RGB}{110,110,110}
\definecolor{adjc}{RGB}{196,30,40}
\definecolor{botc}{RGB}{25,75,190}
\newcommand{\cN}[1]{{\color{normc}#1}}
\newcommand{\cA}[1]{{\color{adjc}#1}}
\newcommand{\cB}[1]{{\color{botc}#1}}

\newcommand{\dd}{\,\mathrm d}
\newcommand{\cusp}{\Gamma_{\mathrm{cusp}}}
\newcommand{\coll}{\Gamma_{\mathrm{coll}}}
\newcommand{\B}{\mathcal B}

\newcommand{\ee}{\mathrm e}

\begin{document}
\title{Large-Spin Structure Constants of $\mathcal N=4$ SYM at Any Coupling}
\author{Alessandro Georgoudis\,\orcidlink{0000-0002-1601-4218}}
\email{alessandro.georgoudis@city.ac.uk}
\affiliation{Department of Mathematics, City St George’s,
University of London, Northampton Square, EC1V 0HB, London, UK}

\begin{abstract}
We compute the structure constant of twist-two operators of large spin
with two half-BPS operators in planar $\mathcal N=4$ super Yang--Mills
theory at any coupling. Crossing symmetry fixes them up to one function
of the coupling, the hard function $H(g)$, which we obtain by taking
the large-spin limit inside the hexagon form factors. The result is a
closed formula built from the BES determinant, an octagon constant and
a coefficient of the scalar flux-tube measure. Its weak-coupling expansion reproduces the three known loops of $H$ and
gives any higher one while at strong coupling we show that it decays exponentially. Surprisingly, the same treatment, applied to the asymptotic formula for
two maximal giant gravitons, gives also the structure constant at any coupling, showing some universality of our representation of large-spinning operators. 
\end{abstract}
\maketitle

\section{Introduction}
Recent advances in the study of structure constants of planar
$\mathcal N=4$ super Yang--Mills
theory~\cite{Basso:2015zoa,Basso:2015eqa,Basso:2017muf,Basso:2022nny,Basso:2025mca}
have led to the first non-perturbative results for single-trace
operators purely from
integrability~\cite{Ekhammar:2026sli}. While low-lying states are now
accessible, here we treat the opposite regime, twist-two operators of
large spin $S$, dual to long folded strings spinning in
AdS~\cite{Maldacena:1997re,Gubser:2002tv}. These operators control the
lightlike limit of four-point
functions~\cite{Alday:2013cwa,Alday:2015eya}, which is related to null
Wilson loops and scattering
amplitudes~\cite{Alday:2010zy,Eden:2010zz}. Especially interesting are
the structure constants of the leading twist-two operator
$\mathcal O_S=\operatorname{Tr}(D^SZ^2)+\cdots$ with two
half-BPS operators $\mathcal O_2$ of dimension two. At leading power in
$1/S$, crossing symmetry fixes them
to~\cite{Alday:2013cwa,Alday:2015eya,Dempsey:2025yiv}
\begin{equation}
 \left(\frac{C_S}{C_S^{\rm tree}}\right)^2
 =H(g)\,2^{-\gamma_S}
 \Gamma\left(1-\frac{\gamma_S}{2}\right)^2
 \ee^{-\gamma_E\gamma_S-\coll L}.
 \label{eq:AB}
\end{equation}
Here $g=\sqrt{\lambda}/(4\pi)$, $L=\log S+\gamma_E$, and
$\gamma_S=2\cusp L+\coll$ is the anomalous dimension of
$\mathcal O_S$, with $\cusp$ and $\coll$ the cusp and collinear
anomalous dimensions, both known from
integrability~\cite{Beisert:2006ez,Basso:2007wd,Freyhult:2009my,Fioravanti:2009xt}.
All remaining dynamics sits in one function of the coupling, the hard
function $H$. 

The same function sets the height of the Sudakov peak of the
energy--energy correlator (EEC) in the back-to-back
limit~\cite{Belitsky:2013ofa,Korchemsky:2019nzm,Dempsey:2025yiv}, which
makes it a collider observable and a target of the conformal
bootstrap~\cite{Dempsey:2025yiv}. In QCD, the part of highest
transcendental weight of the analogous hard function is expected to
coincide with $H$~\cite{Kotikov:2002ab,Belitsky:2013ofa}. So far $H$ is
known through three loops~\cite{Eden:2012rr}.

Integrability should give $H$ exactly. The recent results were
obtained with hexagon form
factors~\cite{Basso:2015zoa,Basso:2015eqa,Basso:2017muf,Basso:2022nny,Basso:2025mca},
which express $C_S$ at any coupling in terms of the spectral data of
$\mathcal O_S$, and with the quantum spectral curve
(QSC)~\cite{Gromov:2014caa}, which supplies these data. At large spin
neither is easy to use. The QSC can be solved numerically spin by
spin, but its large-spin limit is not known in closed form. In the
hexagons, $\mathcal O_S$ carries $S$ Bethe roots, the sum runs over
the $2^S$ ways of sharing them, and the simple form~\eqref{eq:AB}
hides in cancellations among exponentially many terms.

In this Letter we take the large-spin limit inside the hexagons. The
Bethe roots, the zeros of the Baxter Q-function, then become dense and
are described by a smooth density, fixed at any coupling by the
Beisert--Eden--Staudacher (BES) equation~\cite{Beisert:2006ez}, and
the sum over the ways of sharing them can be carried out. Of the
virtual particles that glue the two hexagons together, those next to
$\mathcal O_S$ are suppressed by powers of the spin, and a single
series of them remains, as in Coronado's
octagon~\cite{Coronado:2018ypq}. The density is then the only
spectral data the hexagons need.

To further simplify our result we use a relation between the Adjacent
and Bottom factors of the hexagon representation, found at spin two in
Ref.~\cite{Ekhammar:2026sli}, which we extend to all spins. We then
arrive at a closed formula for $H$ at any coupling. It combines three
different integrability objects, the determinant of the BES kernel,
the constant of the null octagon and a coefficient of the scalar
flux-tube measure. Its weak-coupling expansion gives $H$ to any loop
order: it reproduces the three known loops, and we write out the next
two. At finite coupling we evaluate the formula numerically, and at
strong coupling we find that $H$ decays exponentially. We conclude by
showing that the same treatment of the Bethe roots, applied to the
large-spin operator with two maximal giant gravitons, gives their
structure constant at any coupling~\cite{He:2026onf}.

\begin{figure}[t]
\centering
\begin{tikzpicture}[line cap=round,line join=round,
 every node/.style={font=\footnotesize}]
 \def\R{1.72}
 \coordinate (v0) at (0:\R);   \coordinate (v1) at (60:\R);
 \coordinate (v2) at (120:\R); \coordinate (v3) at (180:\R);
 \coordinate (v4) at (240:\R); \coordinate (v5) at (300:\R);
 \path[drop shadow={opacity=0.18,shadow xshift=0.6mm,shadow yshift=-0.6mm},
  top color=white,bottom color=botc!9]
  (v0)--(v1)--(v2)--(v3)--(v4)--(v5)--cycle;
 \draw[line width=1.8pt] (v2)--(v1);
 \draw[line width=1.8pt] (v0)--(v5);
 \draw[line width=1.8pt] (v3)--(v4);
 \draw[adjc,line width=1.3pt,dash pattern=on 3pt off 2pt] (v1)--(v0);
 \draw[adjc,line width=1.3pt,dash pattern=on 3pt off 2pt] (v3)--(v2);
 \foreach \e/\f in {v3/v2,v0/v1}{
  \coordinate (n) at ($(\e)!0.38!(\f)!0.32cm!90:(\f)$);
  \coordinate (p) at ($(\e)!0.38!(\f)!0.32cm!-90:(\f)$);
  \draw[adjc!35,line width=0.8pt,decorate,decoration={snake,amplitude=1pt,segment length=4pt}] (n)--(p);}
 \draw[botc,line width=1.8pt] (v5)--(v4);
 \foreach \x in {-0.62,-0.31,0,0.31,0.62}
  \draw[botc,line width=0.9pt,decorate,decoration={snake,amplitude=1.3pt,segment length=4.2pt}]
   (\x,-1.95)--(\x,-1.08);
 \fill[normc!22,rounded corners=2pt] (-0.98,1.40) rectangle (0.98,1.63);
 \foreach \x/\s in {-0.86/1,-0.70/0,-0.55/1,-0.42/0,-0.30/1,-0.19/0,-0.09/1,0/0,0.09/1,0.19/0,0.30/1,0.42/0,0.55/1,0.70/0,0.86/1}{
  \ifnum\s=1 \fill[adjc] (\x,1.515) circle (0.034);
  \else \draw[adjc,line width=0.55pt,fill=white] (\x,1.515) circle (0.034);\fi}
 \node at (0,2.02) {$\mathcal O_S$};
 \node[normc,align=center,anchor=west] at (1.08,2.08) {norm $\mathcal N$:\\[-1pt] $S$ roots $\to$ sea};
 \node[adjc,align=center,anchor=east] at (-1.08,2.08) {partitions\\[-1pt] $\alpha\,|\,\bar\alpha$};
 \node at ($(v3)!0.5!(v4)+(-0.45,-0.05)$) {$\mathcal O_{2}$};
 \node at ($(v0)!0.5!(v5)+(0.45,-0.05)$) {$\mathcal O_{2}$};
 \node[adjc,align=center] at (-2.45,0.15) {adjacent\\[-1pt] $\propto S^{-a}$};
 \node[adjc,align=center] at (2.52,0.15) {adjacent\\[-1pt] $\propto S^{-a}$};
 \node[botc] at (0,-2.25) {bottom mirror gas $\B$};
\end{tikzpicture}
\caption{One of the two hexagons of $C_S$ at large spin, with the colour
code used throughout: norm in grey, partitions and adjacent bridges in
red, bottom bridge in blue. The $S$ roots of $\mathcal O_S$ condense into
a sea and are shared, $\alpha\,|\,\bar\alpha$, between the two hexagons.
Mirror particles on the adjacent bridges are suppressed by powers of
$1/S$; those on the bottom bridge survive.}
\label{fig:hexagons}
\end{figure}
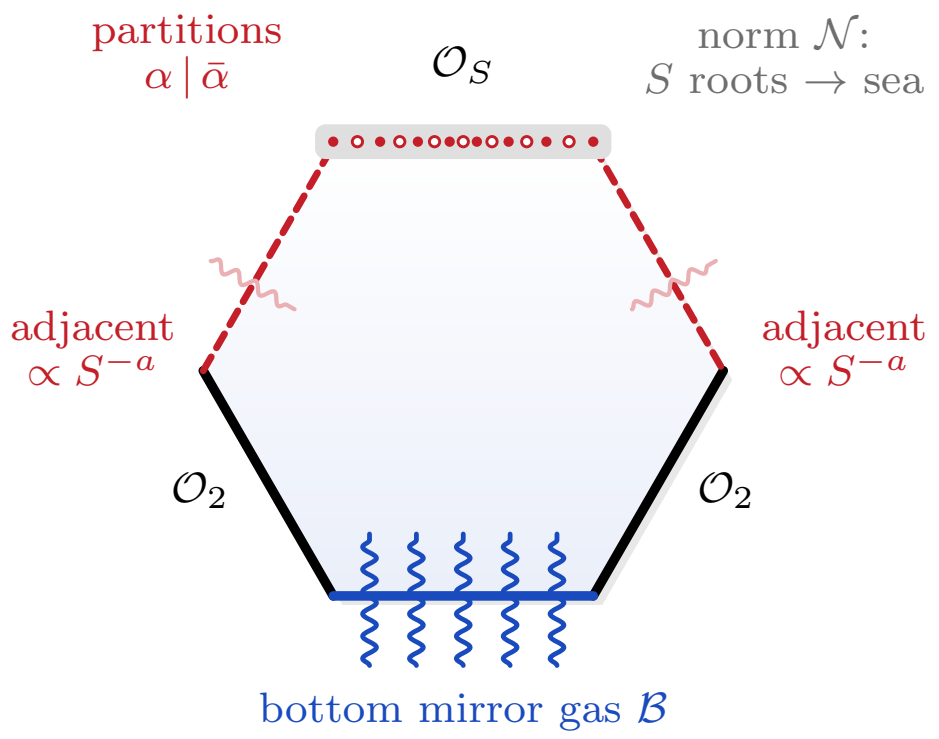

\section{Large Spin Hexagons}
The hexagon approach cuts the three-point function along its three
seams, in our case all of length one. This gives two hexagons
(Fig.~\ref{fig:hexagons}), each a form factor for the excitations on
its three physical edges. To glue
them back we sum over all states that can cross each seam. These are
virtual particles, also called mirror particles. The result
is
\begin{equation}
 \frac{C_S}{C_S^{\rm tree}}
 =\cN{\widehat{\mathcal N}}\,\cA{\widehat{\mathcal A}}\,\cB{\B} ,
 \qquad
 \cN{\widehat{\mathcal N}=\frac{\mathcal N}{\mathcal N^{\rm tree}}},
 \quad
 \cA{\widehat{\mathcal A}=\frac{\mathcal A}{\mathcal A^{\rm tree}}}.
 \label{eq:hexagon}
\end{equation}
At weak coupling the $S$ derivatives of $\mathcal O_S$ are magnons on
the vacuum $\operatorname{Tr}Z^2$, with the Bethe roots as rapidities.
In this picture $\cN{\mathcal N}$ is the norm of the state.
The factor $\cA{\mathcal A}$ sums over the ways of sharing the magnons
between the two hexagons, and includes the virtual particles on the two
bridges adjacent to $\mathcal O_S$. The factor $\cB{\B}$
collects the virtual particles on the bridge facing $\mathcal O_S$,
the bottom bridge, and equals one at tree level. The weights of the
virtual particles are fixed by the hexagon form factors. We now take
$S\to\infty$ at fixed coupling, keeping the leading power in $1/S$
with its full dependence on $L$ and study what happens to the different virtual excitations.

\emph{Adjacent bridges.---}Virtual particles on the adjacent bridges
wind around the spinning operator. Their weights are controlled by the
middle-node functions $Y_{a,0}$~\cite{Basso:2022nny}, which decrease as
powers of $1/S$. At each order in $g$ this suppression survives the
mirror integrals and the sums over bound states, and we assume that is also does at finite coupling. So these particles drop out, together with the
corresponding corrections to the norm and the Bethe roots, just as
wrapping disappears from the twist-two spectrum at large
spin~\cite{Bajnok:2008qj,Lukowski:2009ce}. The norm reduces to its
asymptotic form, the Gaudin norm, and $\cA{\mathcal A}$ to the sum over
partitions without virtual particles.

\emph{Bottom bridge.---}A particle on the bottom bridge scatters with
all $S$ roots, and its weight is proportional to the compact transfer
matrix of $\mathcal O_S$. This factor grows with the spin and
compensates the cost of crossing the bridge, so the bottom series
survives. The transfer matrix has a wrapping correction of its own,
which first appears at five loops and is also suppressed at large spin.
We are left with a structure similar to the one of the
octagon~\cite{Coronado:2018ypq,Belitsky:2019fan}, two hexagons glued
along a single bridge, except that the edge opposite that bridge now
carries the spinning operator.

\emph{The sea.---}The final step is to understand what happens to the spinning operator. At large spin the roots become dense, and sums over
them turn into integrals with a smooth density $\rho(u)$ of roots at
rapidity $u$. This sea of roots is the Gubser--Klebanov--Polyakov (GKP)
state, the long folded
string~\cite{Gubser:2002tv,Belitsky:2006en,Beisert:2006ez}. In Fourier space,
where functions of the Zhukovsky variable $x$, with $x+1/x=u/g$, become
Bessel functions $J_n(2gt)$, the density takes the form
\begin{gather}
 \rho(u)=\frac2\pi\Big[L-\int_0^\infty\frac{\dd t}{\ee^t-1}
 \big(1-\ee^{t/2}\cos(ut)\,\mathcal J(t)\big)\Big],\nonumber\\
 \mathcal J(t)=J_0(2gt)-\sum_{n\ge1}n\,y_nJ_n(2gt).
 \label{eq:rho}
\end{gather}
The Bethe equations fix the coefficients $y_n$, one for each Bessel
mode. They are linear in $L$, $y=Lc+v$, where the vectors $c$ and $v$
solve the BES equations
\begin{gather}
 (1+K)\,c=2g\,e_1,\qquad (1+K)\,v=s,\nonumber\\
 K_{nm}=2m(-1)^{nm+m}\int_0^\infty\frac{\dd t}{t}\,
 \frac{J_n(2gt)J_m(2gt)}{\ee^t-1},\nonumber\\
 s_n=2\int_0^\infty\frac{\dd t}{t(\ee^t-1)}
 \big[J_n(2gt)J_0(2gt)-gt\,\delta_{n1}\big],
 \label{eq:BES}
\end{gather}
for $n,m\ge1$, with $(e_1)_n=\delta_{n1}$. Here $K$ is the interaction
of the roots in the sea, mode by mode. The anomalous dimensions are single components of the solution, $\cusp=2gc_1=4g^2[(1+K)^{-1}]_{11}$ and $\coll=4gv_1$.

\subsection{Norm, partitions and bottom}
Evaluating the three factors of Eq.~\eqref{eq:hexagon} on the sea and
normalizing to tree level, we obtain
\begin{align}
 \cN{\log\widehat{\mathcal N}}&
 \cN{{}=2E(L)-\tfrac32\gamma_S\log2+n_0},\nonumber\\
 \cA{\log\widehat{\mathcal A}}&
 \cA{{}=-E(L)+\gamma_S\log2+a_0},\nonumber\\
 \cB{\log\B}&\cB{{}=-E(L)-\tfrac12(\gamma_E\gamma_S+\coll L)+\log\Gamma(1-\tfrac{\gamma_S}2)+b_0}.
 \label{eq:LS}
\end{align}

Here $E(L)$ is the interaction energy of the density with itself. In
the norm and the partition sum it comes from the pairwise interactions
of the roots, and in the bottom from the scattering of its particles
with the sea. The powers of two come from the largest roots, at
rapidities of order $S$~\cite{Belitsky:2006en}. There the roots behave
as at tree level with the spin shifted to
$S+\frac{\gamma_S}2$~\cite{Basso:2006nk}.

The remaining terms $\cN{n_0}$, $\cA{a_0}$ and $\cB{b_0}$ are
constants, independent of the spin. The norm gives the determinant of
the BES kernel, $\cN{n_0=-\log\det(1+K)}$, and the partition sum the
constant of the null octagon~\cite{Basso:2020xts,Belitsky:2019fan},
$\cA{a_0=-\frac18\log[\sinh(4\pi g)/4\pi g]}$. The constant term of
$\log(\cN{\widehat{\mathcal N}}\cA{\widehat{\mathcal A}})$, extracted
from solutions of the asymptotic Bethe equations at finite spin, agrees
with Eq.~\eqref{eq:LS} through three loops and at finite coupling.

The bottom sum produces, order by order in $g$ the Gamma
function of Eq.~\eqref{eq:AB}, whose poles at $\gamma_S=2n$ mark the
points where the twist of $\mathcal O_S$ meets that of the double
traces $[\mathcal O_2\mathcal O_2]_{n-1,S}$. Its constant $\cB{b_0}$ is
a priori a generic determinant.

\section{Hard function}
When we multiply the three factors, $E(L)$ cancels, and the Alday-Bissi
form~\eqref{eq:AB} emerges out of the three-point function, with $H$
collecting the three constants.

To express $\cB{b_0}$ through known data we use the relation between
adjacent and bottom factors,  which we extend to
all spins as
\begin{equation}
 \cA{\mathcal A}=g^2|\hat\alpha_S|\,(S-1)\,
 \frac{\Gamma(S-1+\frac{\gamma_S}2)}{\Gamma(1-\frac{\gamma_S}2)}\,\cB{\B} \,.
 \label{eq:ABS}
\end{equation}
For $S=2$ is Eq.~(8) of \cite{Ekhammar:2026sli}, with $\cB{\B}$ their
$\mathcal B_2$, and reproduces the known three-loop adjacent
factor~\cite{Basso:2025mca} at spins four and six.

At large spin Eq.~\eqref{eq:ABS} gives
\begin{equation}
 \cB{b_0}=\cA{a_0}-\cB{c_{\rm glue}},\qquad
 \cB{c_{\rm glue}=\frac{\pi^2}{4}\cusp-2\Gamma_{\rm third}}.
 \label{eq:b0}
\end{equation}
Here $\Gamma_{\rm third}$ is the constant term of the scalar flux-tube
measure at large rapidity,
$\log[\mu_\phi(u)/\mu_\phi^{(0)}(u)]=-\frac12\cusp\ell_u^2-\frac12\coll\ell_u
+\Gamma_{\rm third}$ with $\ell_u=\log u+\gamma_E$, given at any
coupling by Eqs.~(5.33)--(5.35) of Ref.~\cite{Basso:2024hlx}, in whose
notation $\Gamma_{\rm virtual}=\coll/2$.

Both sides of Eq.~\eqref{eq:b0} can be computed independently, which
tests Eq.~\eqref{eq:ABS} at large spin. Order by order in $g$,
$\cB{b_0}$ follows directly from the bottom mirror sum and agrees with
the right-hand side through five loops; at four loops it also agrees
with the large-spin limit of the complete finite-spin bottom
integrand~\cite{Basso:2017muf,Basso:2025mca}. At finite coupling, the
large-spin limit of $|\hat\alpha_S|$ obtained from QSC solutions
reproduces $\cB{c_{\rm glue}}$. With Eq.~\eqref{eq:b0}, the three
constants give
\begin{equation}
 \boxed{\begin{aligned}
 H(g)={}&
 \frac{\cB{\exp[\,4\Gamma_{\rm third}(g)-\tfrac{\pi^2}{2}\cusp(g)\,]}}
 {\cN{\det[1+K(g)]^2}}\\[-1pt]
 &\times\cA{\sqrt{\frac{4\pi g}{\sinh(4\pi g)}}} .
 \end{aligned}}
 \label{eq:Hmaster}
\end{equation}
The last factor is $H_{\rm oct}^{\,4}$, where
$H_{\rm oct}=[4\pi g/\sinh(4\pi g)]^{1/8}$ is the constant of one null
octagon~\cite{Belitsky:2019fan}. The octagon and the spinning state are governed by the same BES kernel at two values of its tilt~\cite{Basso:2020xts}, $\alpha=0$ and
$\alpha=\pi/4$.

\emph{Weak coupling.---}
Every ingredient of Eq.~\eqref{eq:Hmaster} is a power series in $g^2$,
so $H$ follows to any loop order. Through five loops,
\begin{align}
 \log H={}&-4\zeta_2g^2+60\zeta_4g^4
 -\left(\tfrac{2146}{3}\zeta_6+\tfrac{272}{3}\zeta_3^2\right)g^6
 \nonumber\\
 &+\left(8916\zeta_8+1360\zeta_3\zeta_5
       +80\zeta_2\zeta_3^2\right)g^8\nonumber\\
 &-\left(\tfrac{580312}{5}\zeta_{10}
       +12768\zeta_3\zeta_7+\tfrac{28128}{5}\zeta_5^2\right.\nonumber\\
 &\hspace{1.2cm}\left.
       +768\zeta_4\zeta_3^2+896\zeta_2\zeta_3\zeta_5\right)g^{10}
 +O(g^{12}).
 \label{eq:logH}
\end{align}
The first three coefficients reproduce the known hard
function~\cite{Eden:2012rr,Alday:2013cwa}, and the next two are
predictions.

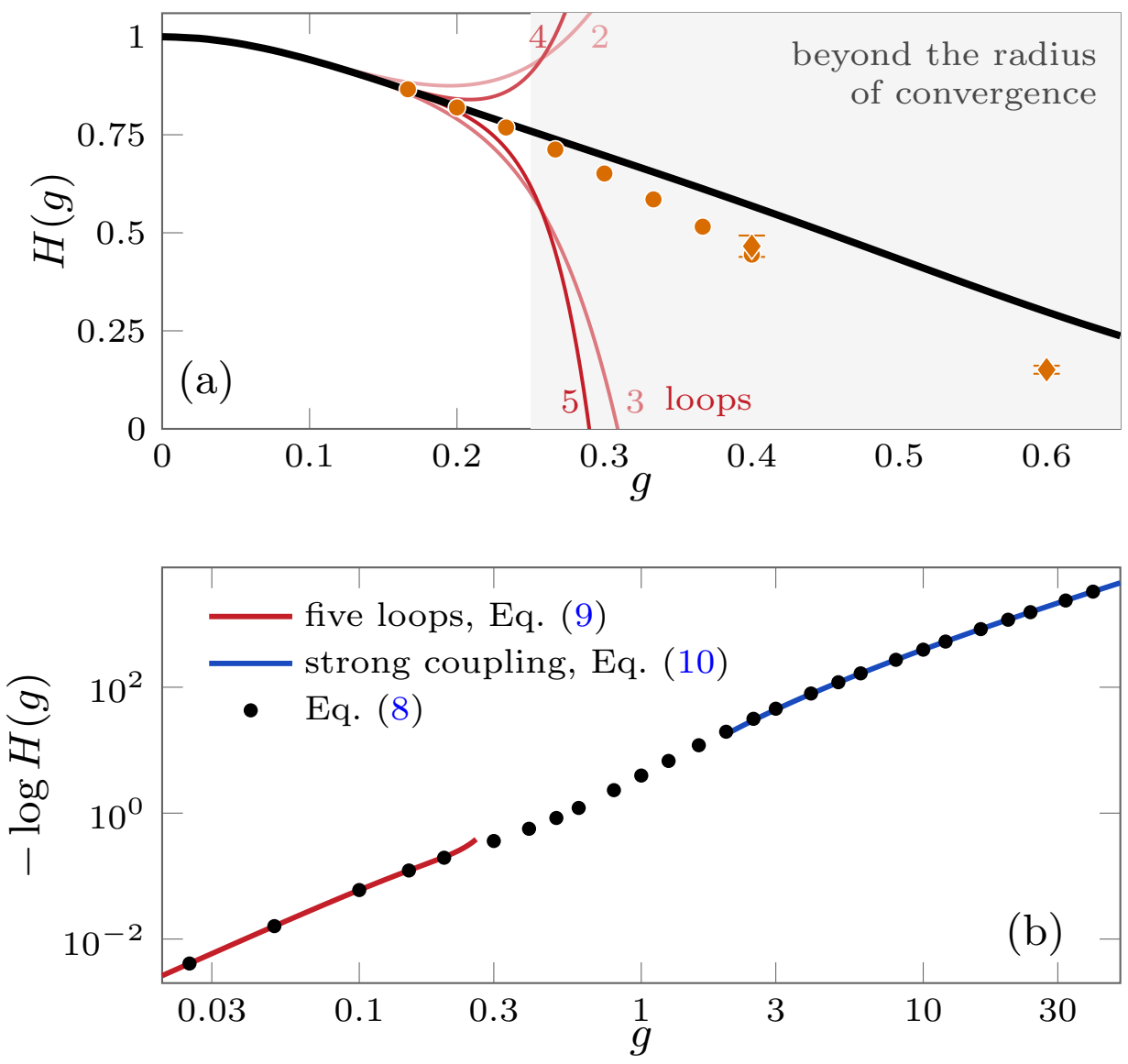
\begin{figure}[t]
\centering
\begin{tikzpicture}
\begin{axis}[name=top,scale only axis,width=0.80\columnwidth,height=3.0cm,
 xmin=0,xmax=0.65,ymin=0,ymax=1.06,restrict y to domain=-1:2,
 axis line style={black!60},tick style={black!60},
 xlabel={$g$},ylabel={$H(g)$},
 xlabel style={yshift=5pt},ylabel style={yshift=-6pt},
 tick label style={font=\scriptsize},label style={font=\small},
 xtick={0,0.1,0.2,0.3,0.4,0.5,0.6},ytick={0,0.25,0.5,0.75,1},
 yticklabels={$0$,$0.25$,$0.5$,$0.75$,$1$},clip=true]
\fill[gray!8] (axis cs:0.25,0) rectangle (axis cs:0.65,1.06);
\node[font=\scriptsize,gray!55!black,anchor=north east,align=right] at (axis cs:0.645,1.03)
 {beyond the radius\\ of convergence};
\addplot[domain=0:0.65,samples=260,adjc!40,line width=0.75pt]
 {1-6.5797362674*x^2+86.5858586969*x^4};
\addplot[domain=0:0.65,samples=260,adjc!60,line width=0.75pt]
 {1-6.5797362674*x^2+86.5858586969*x^4-1333.5074649903*x^6};
\addplot[domain=0:0.65,samples=260,adjc!80,line width=0.75pt]
 {1-6.5797362674*x^2+86.5858586969*x^4-1333.5074649903*x^6+20080.3639444638*x^8};
\addplot[domain=0:0.65,samples=260,adjc,line width=0.75pt]
 {1-6.5797362674*x^2+86.5858586969*x^4-1333.5074649903*x^6+20080.3639444638*x^8-303464.7363678330*x^10};
\node[font=\scriptsize,adjc!80] at (axis cs:0.255,1.0) {4};
\node[font=\scriptsize,adjc!45] at (axis cs:0.297,1.0) {2};
\node[font=\scriptsize,adjc] at (axis cs:0.277,0.07) {5};
\node[font=\scriptsize,adjc!60] at (axis cs:0.321,0.07) {3};
\node[font=\scriptsize,adjc,anchor=west] at (axis cs:0.330,0.07) {loops};
\addplot[black,line width=1.5pt,smooth]
 coordinates {(0.000,1.00000)(0.025,0.99592)(0.050,0.98407)(0.075,0.96551)(0.100,0.94170)(0.125,0.91421)(0.150,0.88444)(0.175,0.85349)(0.200,0.82208)(0.225,0.79064)(0.250,0.75932)(0.275,0.72810)(0.300,0.69687)(0.325,0.66545)(0.350,0.63368)(0.375,0.60143)(0.400,0.56862)(0.425,0.53525)(0.450,0.50138)(0.475,0.46713)(0.500,0.43270)(0.525,0.39832)(0.550,0.36427)(0.575,0.33085)(0.600,0.29834)(0.625,0.26704)(0.650,0.23721)};
\addplot[only marks,mark=*,mark size=1.8pt,orange!85!black,
 mark options={draw=white,line width=0.3pt}]
 coordinates {(0.1667,0.8660)(0.2000,0.8196)(0.2333,0.7685)(0.2667,0.7124)(0.3000,0.6512)(0.3333,0.5852)(0.3667,0.5156)(0.4000,0.4443)};
\addplot[only marks,mark=diamond*,mark size=2.7pt,orange!85!black,
 mark options={draw=white,line width=0.3pt},
 error bars/.cd,y dir=both,y explicit,
 error bar style={orange!85!black,line width=0.5pt}]
 coordinates {(0.4000,0.4658) +- (0,0.0272) (0.6000,0.1508) +- (0,0.0102)};
\node[font=\small] at (axis cs:0.03,0.12) {(a)};
\end{axis}
\begin{axis}[at={(top.south)},anchor=north,yshift=-1.0cm,
 scale only axis,width=0.80\columnwidth,height=3.0cm,
 xmode=log,ymode=log,xmin=0.02,xmax=50,ymin=2e-3,ymax=8e3,
 axis line style={black!60},tick style={black!60},
 xlabel={$g$},ylabel={$-\log H(g)$},
 xlabel style={yshift=5pt},ylabel style={yshift=-4pt},
 tick label style={font=\scriptsize},label style={font=\small},
 xtick={0.03,0.1,0.3,1,3,10,30},xticklabels={$0.03$,$0.1$,$0.3$,$1$,$3$,$10$,$30$},
 ytick={1e-2,1,1e2,1e4},
 legend style={font=\scriptsize,at={(0.03,0.97)},anchor=north west,draw=none,fill=none,row sep=-1pt},
 legend cell align=left]
\addplot[domain=0.02:0.26,samples=80,adjc,line width=1.1pt]
 {6.5797362674*x^2-64.9393940227*x^4+858.7473693975*x^6-10837.6670813358*x^8+140740.7499568118*x^10};
\addlegendentry{five loops, Eq.~(\ref{eq:logH})}
\addplot[domain=2:50,samples=120,botc,line width=1.1pt]
 {4*x*(ln(x)+0.8840687)^2-1.0044257*x};
\addlegendentry{strong coupling, Eq.~(\ref{eq:strong})}
\addplot[only marks,mark=*,mark size=1.25pt,black]
 coordinates {(0.025,0.0040872)(0.05,0.016056)(0.1,0.060066)(0.15,0.1228)(0.2,0.19592)(0.3,0.36116)(0.4,0.56454)(0.5,0.83772)(0.6,1.2095)(0.8,2.3178)(1,3.9567)(1.25,6.752)(1.6,11.976)(2,19.645)(2.5,31.495)(3,45.573)(4,79.498)(5,120.01)(6,166.15)(8,272.71)(10,395.29)(12,531.38)(16,837.42)(20,1181.1)(24,1556.3)(32,2384.9)(40,3299.9)};
\addlegendentry{Eq.~(\ref{eq:Hmaster})}
\node[font=\small] at (axis cs:25,0.012) {(b)};
\end{axis}
\end{tikzpicture}
\caption{The hard function at finite coupling. (a) Weak coupling:
Eq.~\eqref{eq:Hmaster} (black) against the weak-coupling series of $H$,
the expansion of the exponential of Eq.~\eqref{eq:logH}, truncated at
two to five loops (red), which converges for $g<\frac14$. The points
are the estimates of $H$ obtained in Ref.~\cite{Dempsey:2025yiv} from
the EEC, through an inversion formula (circles) and from fits to
bootstrap bounds on its back-to-back limit (diamonds), read off from
Figs.~17(b) and 15(b) of that reference. (b) From weak to strong coupling:
Eq.~\eqref{eq:Hmaster} (dots) against the five-loop series and the
strong-coupling asymptotics~\eqref{eq:strong}.}
\label{fig:H}
\end{figure}

\emph{Finite coupling.---}We evaluate Eq.~\eqref{eq:Hmaster} at any
coupling by truncating $K$ to finitely many Bessel modes, and all
values quoted here are stable under increasing their number. Beyond the
radius $g=\frac14$ of the perturbative series~\eqref{eq:logH}, $H$
keeps decreasing monotonically (Fig.~\ref{fig:H}a), from
$H(0.4)\simeq0.569$ to $H(0.8)\simeq0.098$ and $H(4)\simeq3\cdot10^{-35}$.

\emph{Strong coupling.---}At large $g$ we find
\begin{equation}
 \log H=-4g\Big(\log\frac g2+\gamma_E+1\Big)^2+\Big(12-\frac{7\pi}2\Big)g+O(\log^2g),
 \label{eq:strong}
\end{equation}
which the formula follows from $g\approx2$ on (Fig.~\ref{fig:H}b). Its
leading term is $-\coll^2/(2\cusp)$, with
$\coll\simeq-4g[\log(g/2)+\gamma_E+1]$ and $\cusp\simeq2g$, so $H$
decays exponentially. 

Equation~\eqref{eq:strong} takes $S\to\infty$ before $g\to\infty$. At
fixed $S/g$ the structure constants are instead those of the classical
folded string~\cite{Jiang:2016ulr,Basso:2025mca}. With
$\Lambda=\log(2S/g)$, the substitution $L=\Lambda+\log(g/2)+\gamma_E$
carries the logarithms of spin from one regime to the other, but not the
constant terms, which depend on the order of the limits.

\emph{Energy correlator.---}The relation between $H$ and the
back-to-back EEC neglects operators of higher twist. This is justified
at weak coupling but not at strong coupling, where level crossing makes
them contribute~\cite{Korchemsky:2019nzm}, so we do not expect
Eq.~\eqref{eq:strong} to describe the EEC in that regime. We believe
that the relation fails already at moderate coupling, because the spins
that dominate the back-to-back EEC, those with $\gamma_S\simeq1$,
decrease quickly, from $S\sim10^5$ at $g=0.1$ to $S\simeq20$ at $g=0.2$
and $S\simeq2$ at $g=0.4$. The values of $H$ that
Ref.~\cite{Dempsey:2025yiv} extracts from the EEC agree with
Eq.~\eqref{eq:Hmaster} for $g\lesssim0.2$ and fall below it at larger
coupling (Fig.~\ref{fig:H}a).

\section{Conclusion}
In this Letter we computed the hard function at finite coupling by
taking the large-spin limit inside the hexagons. Twist-two operators are
the shortest in the theory, and one might expect wrapping to be a
serious obstacle. At large spin there is none. Virtual particles winding
around the spinning operator are suppressed by powers of the spin, as in
the twist-two spectrum, and we are left with two hexagons glued along a
single bridge. Since the Bethe roots form a sea, the sum over the $2^S$
ways of sharing them turns into a few objects built from the BES kernel.
The form~\eqref{eq:AB}, which crossing symmetry dictates from the
four-point function, then comes out of the three-point function
itself, with the powers of two from the norm and the partitions and the
Gamma function from the bottom.

The hard function also sets the height of the Sudakov peak of the
back-to-back EEC, and for $g\lesssim0.2$ our formula agrees with the
values extracted from this observable in Ref.~\cite{Dempsey:2025yiv}.
Its weak-coupling series, which is given to all loops by \eqref{eq:Hmaster} and explicitly through five loops in \eqref{eq:logH}, is also
expected to give the terms of highest transcendental weight of the hard
function of QCD, by the principle of maximal
transcendentality~\cite{Kotikov:2002ab,Belitsky:2013ofa}.

There are several directions for future work and comments:
\begin{itemize}
\item We find it intriguing that the closed form of $H$ depends on
$\Gamma_{\rm third}$, a coefficient of the scalar flux-tube measure.
This measure was bootstrapped for the pentagon description of null
Wilson loops~\cite{Basso:2013aha} and enters the operator product
expansion of form factors~\cite{Basso:2024hlx}. It would be very
interesting to relate pentagons and hexagons at large spin.
\item Surprisingly substituting maximal giant gravitons in place of the two operators
$\mathcal O_2$ seems to be controlled by similar objects. There the structure constant
$D_S$ of $\mathcal O_S$ scales as $2^{-S}S^{d(g)}$, with an exponent
proposed at any coupling in Ref.~\cite{He:2026onf}. At large spin we find that the
asymptotic formula of that reference becomes
\begin{equation}
 \log\frac{D_S}{D_S^{\rm tree}}=-\gamma_S\log2
 -\int_0^\infty\frac{\dd t}{t}\,\frac{1-\mathcal J(-t)}{\ee^{t/2}+1},
 \label{eq:giant}
\end{equation}
with $\mathcal J$ as in Eq.~\eqref{eq:rho}. Its part linear in $L$
gives the exponent, and the rest the constant that multiplies the
power. No higher power of $L$ appears. The two giant gravitons are
joined by order $N$ propagators, so the bridge between them is
infinitely long and there is no bottom channel, which for $C_S$
produces the Gamma function of Eq.~\eqref{eq:AB}. Together they act on
$\mathcal O_S$ as an integrable boundary, which absorbs the roots in
pairs $(u,-u)$, each pair independently, so that the density of the sea enters only linearly.
\item At strong coupling, the main open question is the interpolation
in $S/g$ between Eq.~\eqref{eq:strong} and the classical folded string,
which share their logarithms of spin but not their constant terms.
\item Beyond this structure constant, we expect the same ingredients, a
sea of roots probed by mirror particles through the BES kernel, to
hold. The simplest cases are longer bridges, which change only the
mirror propagation, and higher twist, which adds holes to the sea.
\item Another interesting direction would be to study several spinning
operators~\cite{Bercini:2022gvs}. This would require a joint treatment of their seas, and at
large spin it should match the null hexagonal Wilson loop, to which
three-point functions of spinning operators are
dual~\cite{Bercini:2020msp,Bercini:2021jti}.
\item Finally in higher-point functions these large-spin factors should be
the building blocks of
hexagonalization~\cite{Fleury:2016ykk,Eden:2016xvg,Coronado:2018ypq}. When reconstructing the four point function they should recover the form obtained in \cite{Alday:2013cwa} by CFT methods.
\end{itemize}
\begin{acknowledgments}
I thank Benjamin Basso, Carlos Bercini, Simon Ekhammar, Vasco Gon\c{c}alves, Nikolay Gromov, Andrea Guerrieri, Shota Komatsu and Congkao Wen for insightful discussions on related topics. I am supported by a Royal Society grant URF$\backslash$R1$\backslash$241371.

\paragraph{Use of AI tools.}
AI tools were used in preparing this work. Symbolic and numerical verifications were performed in Mathematica through the Wolfbook interface~\cite{Wolfbook}. The author takes full responsibility for the content of the paper.
\end{acknowledgments}

\bibliography{prl}
\end{document}